\documentstyle[preprint,aps,psfig]{revtex}
\begin{document}
\draft
\title{Designability and Thermal Stability of Protein Structures}
\author{Ned S. Wingreen$^{\rm a}$, Hao Li$^{\rm b}$, Chao
Tang$^{\rm a}$\thanks{Corresponding author. Tel.: +1-609-951-2644;
fax: +1-609-951-2496. {\it E-mail address}: tang@nec-labs.com(C.Tang).}}
\address{$^{\rm a}$NEC Laboratories America, Inc., 4 Independence Way, 
Princeton, NJ 08540, USA}
\address{$^{\rm b}$Department of Biochemistry and Biophysics, University of
California at San Francisco, San Francisco, CA 94143, USA}

\maketitle

\begin{abstract}
Only about 1,000 qualitatively different protein folds are
believed to exist in nature. Here, we review theoretical studies 
which suggest that some folds are intrinsically more designable 
than others, {\it i.e.} are lowest energy states of an unusually 
large number of sequences. The sequences associated with these 
folds are also found to be unusually thermally stable. The connection 
between highly designable structures and highly stable sequences 
is generally known as the ``designability principle''. 
The designability principle may help explain the small number 
of natural folds, and may also guide the design of new folds.

\end{abstract}

\section{Introduction}

Two remarkable features of natural proteins are the simple fact
that they fold and the limited number of distinct folds they adopt.
Random amino-acid sequences typically do not fold to a 
unique structure. Rather they have many competing configurations
of similar minimum free energy. Nature has evolved sequences that do
fold stably, but it is estimated that the total number of qualitatively
different folds is only about 1,000~\cite{cho-fold,orengo,brenner}.

To attempt to explain these remarkable features of natural 
proteins, we have proposed a principle of 
designability~\cite{helling01,fink87,yue95,gov95,li-sci}. 
The designability of a structure is the number of sequences 
which have that structure as their unique lowest-energy
configuration~\cite{li-sci}. In a wide range of models, structures
are found to differ dramatically in designability, and sequences 
associated with highly designable structures have unusually 
high thermal stability~\cite{li-sci,li-pnas,miller02,li02,eldon02}. 
We refer to this connection between the designability of a structure
and the stability of its associated sequences as the designability 
principle. In model studies, highly designable 
structures are rare. As a result, thermally
stable sequences are also rare. We conjecture that the
designability principle also applies to real proteins,
and that natural protein folds are exceptional, highly 
designable structures.

In this article, we review the designability principle.
We start from a minimal model of protein structure in
which the designability of a structure can be understood
geometrically as the size of its basin of attraction in 
sequence space. More detailed models, including
all 20 amino-acid types and off-lattice backbone configurations,
reinforce the basic principle and provide a framework
for the design of qualitatively new protein folds.

\section{Purely Hydrophobic (PH) Model}

Generally, the folding of proteins relies on the formation
of a hydrophobic core of amino acids. Consideration of 
hydrophobicity alone leads to a very simple description
of proteins--the ``purely hydrophobic'' (PH) model~\cite{li-pnas}. 
In this model, sequences consist of only two types of amino acids, 
hydrophobic and polar~\cite{lau89}. Structures are compact walks 
on a cubic or square 
lattice. An example of a $6\times6$ square structure is shown 
in Fig.~\ref{fig1}(a). As indicated, each structure consists 
of core sites surrounded by surface sites. The energy of a 
particular sequence folded into a particular structure is
the number of hydrophobic amino acids occupying core sites,
multiplied by $-1$, 
\begin{equation}
E=-\sum_{i=1}^N s_i h_i.
\label{ham1}
\end{equation}
A binary string $\{s_i\}$ represents each folded structure:
$s_i=1$ if site $i$ along the chain is in the core,
and $s_i=0$ if the site is on the surface.
Similarly, a binary string $\{h_i\}$ represents each sequence:
$h_i=1$ if the $i$th amino acid in the sequence is hydrophobic, 
and $h_i=0$ if the amino acid is polar.

Within the PH model, structures differ dramatically in their 
designability. In practice, the designability 
is obtained by sampling a large number of binary sequences, 
and, for each sequence, recording the unique lowest-energy 
structure if there is one. Finally, the number of sequences which 
map to, {\it i.e.} ``design", 
each structure is summed to give the designability
$N_S$ of the structure.  Fig.~\ref{fig1}(b) shows a histogram of 
designability $N_S$ for compact $6\times6$ structures.
There are 57,337 structures, with 30,408 distinct binary strings.
Most structures have a designability around 50, but 
a small number of structures have designabilities more
than 10 times this high. If sequences were randomly assigned
to structures, the result would the Poisson distribution
which is shown for comparison, and there would be no structures 
with such high designability.  

Importantly, the PH model has a simple geometrical 
representation that explains both the wide range of designabilities and the close connection between designability and thermal stability. 
To find the relative energies of  
different structures for a given sequence, the energy in 
Eq.~(\ref{ham1}) can be replaced by
\begin{equation}
E = \sum_{i=1}^N (h_i - s_i)^2.
\label{ham2}
\end{equation}
This replacement is allowed because the extra term $\sum h_i^2$ 
is a constant for a given sequence, and the other extra term
$\sum s_i^2$ is also a constant, equal to the number of core sites,
for all compact structures. Eq.~(\ref{ham2}) indicates
that the energy of a sequence folded into a particular
structure is simply the {\it Hamming distance}~\cite{Hamming} 
between their respective binary strings. So, for a given sequence, 
the lowest-energy structure is simply the closest structure. 
The designability of a structure is therefore the exclusive 
volume of binary strings (sequences) that lie closer to it than to
any other structure, as shown schematically in Fig.~\ref{fig2}.

The wide range of structure designabilities can be traced to the 
nonuniform density of structures in the space of binary strings. Most
structures are found in dense regions, {\it i.e.} in clusters of
structures with similar patterns of surface and core sites. Structures
found in these crowded regions have small exclusive volumes, and so, by
definition, have small designabilities. In fact, many groups of distinct
structures share an identical surface-core pattern 
(binary string) and therefore have
zero designability. In contrast, a few structures fall in low-density
regions, that is they have unusual surface-core patterns, and so have 
large exclusive volumes. These are the highly designable structures. 
In Fig.~\ref{fig3}(a), we plot the number of structures $n(d)$ at a
Hamming distance $d$ from a structure with low, intermediate, and high
designability, respectively. It shows that both low- and high-density 
neighborhoods typically have a large spatial extent, reaching nearly halfway 
across the space of binary strings. 

The geometrical representation of the PH model makes clear the 
connection between thermal stability and designability. A sequence is
considered to be thermally unstable if it has a small or vanishing energy
gap between its lowest-energy structure and all other structures, and if
there are many such competing structures. In the PH model, the energy of
a sequence folded into a particular structure is the distance
between their binary strings. A sequence which folds to a structure in
a dense region (cf. Fig.~\ref{fig2}) will necessarily lie close to many
other structures, and will therefore have many competing low-energy 
conformations. Even if the sequence perfectly matches the structure, 
with hydrophobic amino acids at all core sites and polar amino acids 
at all surface sites, the high surrounding density of structures with 
similar surface-core patterns implies a large number of competing folds.
This is the hallmark of thermal instability. Therefore, the low-designability 
structures, which are found in high-density regions, will have associated
sequences which are thermally unstable. 

In contrast, if a sequence folds to a structure in a low-density region,
there will be relatively few nearby structures, and so relatively few
competing folds. These sequences will be thermally stable. Therefore, 
the highly designable structures, from low-density regions, will have
associated sequences of high thermal stability. This is the designability
principle in a nutshell--high designability and thermal stability are
connected because both arise from low-density regions in the space of 
binary strings which represent folded structures.

A measure of the ``neighborhood density'' of structures around a
particular structure is the variance $\gamma$ of the quantity $n(d)$ 
shown in Fig.~\ref{fig3}(a). The variance $\gamma$ is directly related
to the thermal stability--smaller $\gamma$ implies lower neighhood density
and hence higher thermal stability. In Fig.~\ref{fig3}(b) we plot this
variance as a function of designability. It shows a strong correlation
between the designability and the thermal stability.

Since low-energy competing structures also represent kinetic traps,
one expects the thermally stable sequences associated with highly 
designable structures to be fast folders as well. This has been tested
for a lattice model closely related to the PH model~\cite{regis}.

\section{Miyazawa-Jernigan (MJ) Matrix Model}

Natural proteins contain 20 amino acids, not two, and their interactions
are more complicated than simple hydrophobic solvation. Some of these 
real-world features are captured in the Miyazawa-Jernigan (MJ) matrix model.
The MJ matrix is a set of amino-acid interaction energies inferred from 
the propensities of different types of amino acids to be neighbors in 
natural folded structures~\cite{mj85}. The model assigns the appropriate 
energy from the MJ matrix to every pair of amino acids that are on 
neighboring lattice sites, but are not adjacent (covalently bonded) on 
the chain, as indicated in Fig.~\ref{fig4}(a)~\cite{li02}. In studies using
the MJ-matrix model, there are generally too many possible sequences 
(20$^N$) to enumerate, but the relative designabilities of structures 
can be obtained accurately by random sampling of sequences.

Fig.~\ref{fig5}(a) shows a histogram of designability for compact $6\times6$
structures obtained using the MJ matrix of interaction energies. The form
of the distribution is very similar to the PH-model histogram, including 
the tail of highly designable structures (Fig.~\ref{fig1}(b)).  There is
also a strong correlation between thermal stability and designability in the
MJ-matrix model~\cite{li02}.  For thermal
stability, one can use some measure of ``neighborhood" density of states.
We find that in the MJ-matrix model, the
thermal stability of a sequence folded into a structure is well correlated
with the local energy gap~\cite{sali} between the lowest-energy structure
and the next
lowest. Fig.~\ref{fig5}(b) shows the energy gap averaged over sequences 
which fold to structures of a given designability $N_S$. With increasing 
designability, there is a clear increase in the average gap, and hence in 
the thermal stability of associated sequences. The results of the MJ-matrix
model are very similar to those obtained with the PH model. Indeed, the
same structures are found to be highly designable in both models, including
the same top structure shown in Fig.~\ref{fig4}(a). The most designable 
$3\times3\times3$ structure is shown in Fig.~\ref{fig4}(b). Qualitatively, 
the results of the MJ-matrix model are the same for three-dimensional 
structures (Fig.~\ref{fig6}) as for two-dimensional ones.

Why are the results of the purely hydrophobic model and the 
Miyazawa-Jernigan-matrix model so similar? In fact, both models are 
dominated by hydrophobic solvation energies. The interaction energy 
between any two amino acids $i$ and $j$ in the MJ matrix can be well 
approximated by $-(h_i + h_j)$, where $h_i$ is an effective hydrophobicity
for each amino acid~\cite{li97}. This implies that the energy of formation
of a 
non-covalent nearest-neighbor pair is simply the desolvation energy of 
shielding one face of each amino acid from the surrounding water. As a
result, the MJ-matrix model can be viewed as a variant of the PH model
in which there are 20 possible values of hydrophobicity instead of just
two. An additional distinction is that the MJ-matrix model has a range 
of different site types (core, edge, and corner in two dimensions; core,
edge, face, and corner in three dimensions) rather than just surface and
core as in the PH model. Overall, these differences are not enough to 
alter the basic designability principle, or even to change the set of
highly designable $6\times6$ structures. 

For the MJ-matrix model, one can still construct 
a space of structure strings, now including several levels of solvent
exposure between 0 (most exposed) and 1 (most buried). As in the PH model,
some regions of this space are dense with structures and some have few
structures. Structures with similar surface-exposure strings compete
for sequences. As a result, structures in high-density regions have
small basins of attraction for sequences, and structures in low-density
regions have large basins. Moreover, sequences associated with structures
in low-density regions have few competing conformations and are the most
thermally stable. Therefore, the designability principle holds in the 
MJ-matrix model for the same reason it does in the PH model:
high designability and high thermal stability
are connected because both arise in low-density regions in the space of
strings, {\it i.e.} the space of surface-exposure patterns of
structures. 

Lattice-protein models in which hydrophobic solvation does not dominate may
show different behavior. For example, Buchler and Goldstein reported results
of a variant of the MJ-matrix model in which the dominant hydrophobic term
$-(h_i + h_j)$ had been subtracted out~\cite{Buch00}. They found a set of
highly designable structures different from that obtained with the full MJ
matrix, and similar to the set obtained for a random pairing potential
between amino acids.

\section{Off-lattice models}

Natural proteins fold in three dimensions, and their main degrees of freedom
are bond rotations. Does the designability principle extend to off-lattice
models with more realistic degrees of freedom? One model for which 
designability has been studied off-lattice is a 3-state discrete-angle model,
of the type introduced by Park and Levitt~\cite{Park95}. The results 
strongly confirm the designability principle, and suggest the possibility of
creating new, highly designable folds in the laboratory~\cite{miller02}.

The main degrees of freedom of a protein backbone are the dihedral
angles $\phi$ and $\psi$. Certain pairs of $\phi$-$\psi$ angles are preferred
in natural structures, since they lead to conserved secondary structures
such as $\alpha$-helices and $\beta$-strands~\cite{rama}. Discrete-angle
models for
protein structure take advantage of these preferences by allowing only
certain combinations of angles.  For an $m$-pair model, the total number
of backbone structures grows as $m^N$. With $m=3$, it is possible to 
computationally enumerate all structures up to roughly $N=30$ amino acids.

Figure~\ref{fig7}(a) shows an example of a protein backbone of length
$N=23$ generated using a 3-state model with ($\phi$,$\psi$) = (-95,135),
(-75,-25), and (-55,-55), where the first pair of angles corresponds to a 
$\beta$-strand and the second two correspond to variants of $\alpha$-helices.
Structures are decorated with spheres representing sidechains, as shown
in Fig.~\ref{fig7}(b). Only compact self-avoiding structures are 
considered as possible protein folds.

To assess designability of these off-lattice structures, the solvent-exposed
area of each sidechain sphere is evaluated. An energy of hydrophobic solvation
is defined as in Eq.~(\ref{ham1}) by $E=-\sum_{i=1}^N s_i h_i$, where now 
$s_i$ is the fractional exposure to solvent of the $i$th sidechain, and the
$h_i$ are amino-acid hydrophobicities. Figure~\ref{fig8}(a) shows a histogram
of designability for the 3-state model. There is a wide range of 
designabilities, with a tail of very highly designable structures. A strong 
correlation exists between the designability of a structure and the thermal
stability of its associated sequences, as shown in Fig.~\ref{fig8}(b). 

The designability principle evidently applies to the 3-state model 
as well as to the lattice models discussed earlier. This is not surprising, 
because folding in the 3-state model is also driven by hydrophobic 
solvation. Each structure in the model is represented by a string of 
sidechain solvent exposures, represented by real numbers between 0 and 1.
Again, the space of these strings has high- and low-density regions, with
the, by now familiar, relation between low density and high designability 
and thermal stability leading to the designability principle.

A major advantage of the 3-state model is that it addresses
structures that a real polypeptide chain can adopt. Among the highly 
designable folds, one recovers several recognizable natural structures,
including an $\alpha$-turn-$\alpha$ fold and a zincless zinc-finger.
In addition, some of the highly designable folds, including a 
$\beta$-$\alpha$-$\beta$ structure, have not been observed in nature as
independent domains. Results of our effort to create this 
fold in the laboratory are encouraging~\cite{fan}.

\section{Discussion and Conclusion}

The designability principle has been explored in a number of models 
for proteins, including all 20 amino acids and realistic backbone 
conformations. In these models, the strong link between designability 
and thermal stability can be traced to the dominance of the hydrophobic 
solvation energy. Whenever hydrophobicity is dominant, each structure
can be reduced to its pattern of solvent exposure. In the same vein,
each sequence can be reduced to its pattern of hydrophobicity.
Sequences will fold so as to best match their hydrophobic residues 
to the buried sites of structures. Both designability (number of
sequences per structure) and thermal stability depend on a competition
among structures with similar patterns of solvent exposure. Highly 
designable structures are those with unusual patterns of surface
exposure, and therefore with few competitors. This lack of competitors
also implies that the sequences folding to highly designable structures 
are thermally stable.
 
Since hydrophobicity is generally accepted to be the dominant force 
for folding of real proteins, the designability principle may provide 
a guide to understanding the selection of natural protein structures. 
Of course, real proteins are held together by forces other than 
hydrophobicity. Next to hydrophobicity, the formation of hydrogen
bonds is the most important factor in determining how a typical
protein folds. The backbone hydrogen bonds of
$\alpha$-helices and $\beta$-sheets help stabilize particular 
folds. These secondary structures can be incorporated within the  
framework of designability as a favorable energy bias
for formation of $\alpha$-helices and $\beta$-sheets.

One way to incorporate hydrogen bonding in the design of new
protein folds is to specify in advance the secondary structure 
of the protein. This approach has the added advantage of greatly reducing
the number of degrees of freedom. The desired secondary structures can be
designed into the sequence via the propensities of particular
amino acids to form $\alpha$-helices and $\beta$-strands.

This approach to design was recently carried out for four-helix
bundles~\cite{eldon02}. Compact, self-avoiding structures
consisting of four tethered 15-residue $\alpha$-helices were
generated and assessed for designability. Figure~\ref{fig9}
shows the four most designable distinct folds, which closely
correspond to natural four-helix bundles. As shown in 
Fig.~\ref{fig10}, the histogram of designability for the four-helix
model has the characteristic long tail of highly designable
structures.

The principle of designability has been motivated here in
terms of hydrophobic solvation. More generally, the 
dependence of both designability and thermal stability 
on a competition among structures broadens
the application of the principle. For example, designability
and thermal stability have been found to correlate in
non-solvation models including random-interaction models~\cite{Buch00} 
and folding of two-letter RNA~\cite{ranjan}. In the future, we hope 
that designability will provide a guide to the design of new structures
both for polymers other than proteins and for solvents other than water.

We gratefully acknowledge the contributions of many 
coworkers in developing the notion of designability, 
in particular Eldon Emberly, Robert Helling, 
R\'egis M\'elin, Jonathan Miller, 
Tairan Wang, and Chen Zeng.

\newpage

\begin{figure}
\end{figure}
\vspace{0.5cm}

%Fig. 1 (from Review Fig. 12)
\begin{figure}
\centerline{\psfig{file=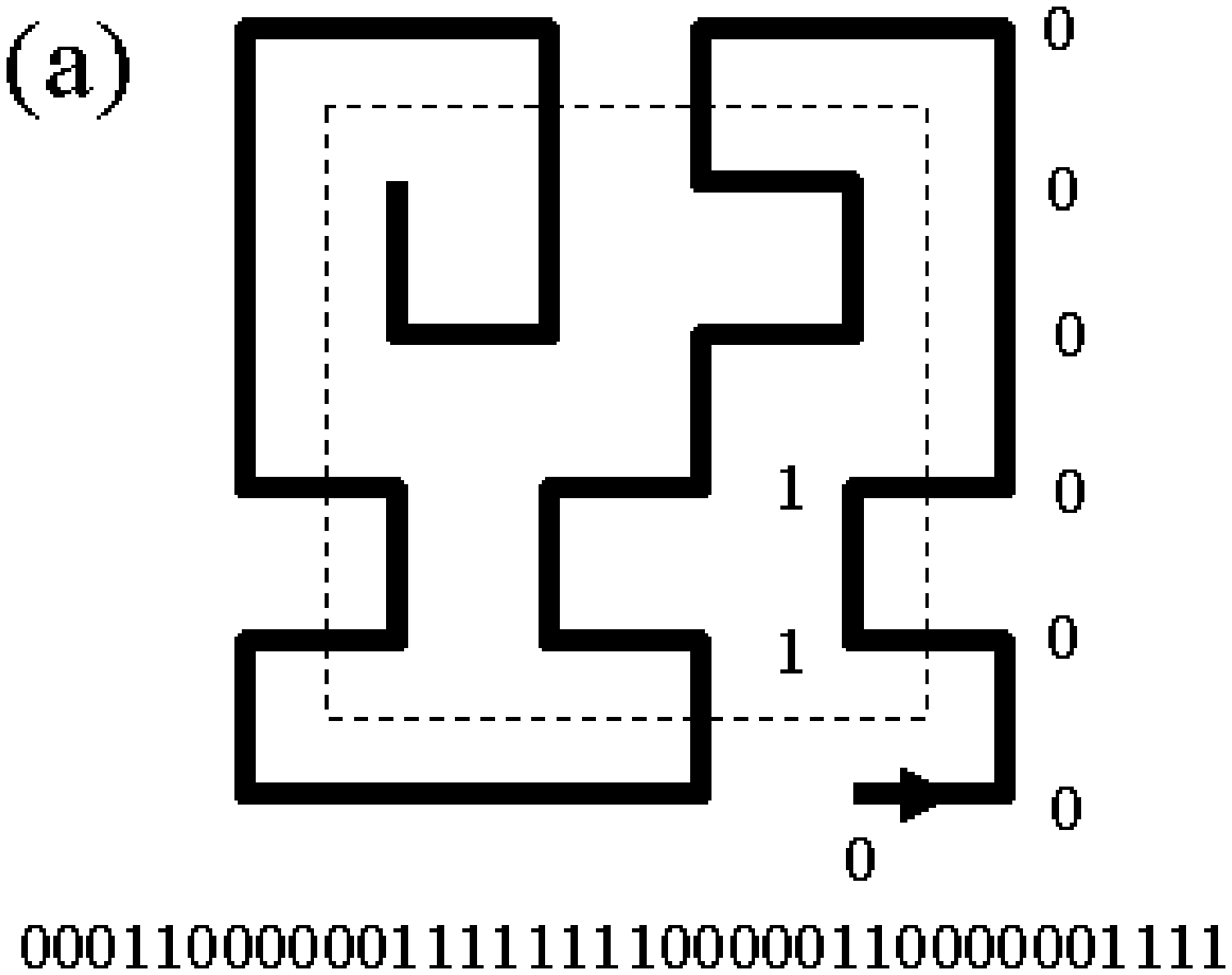,width=3in}}
\vspace{1cm}
\centerline{\psfig{file=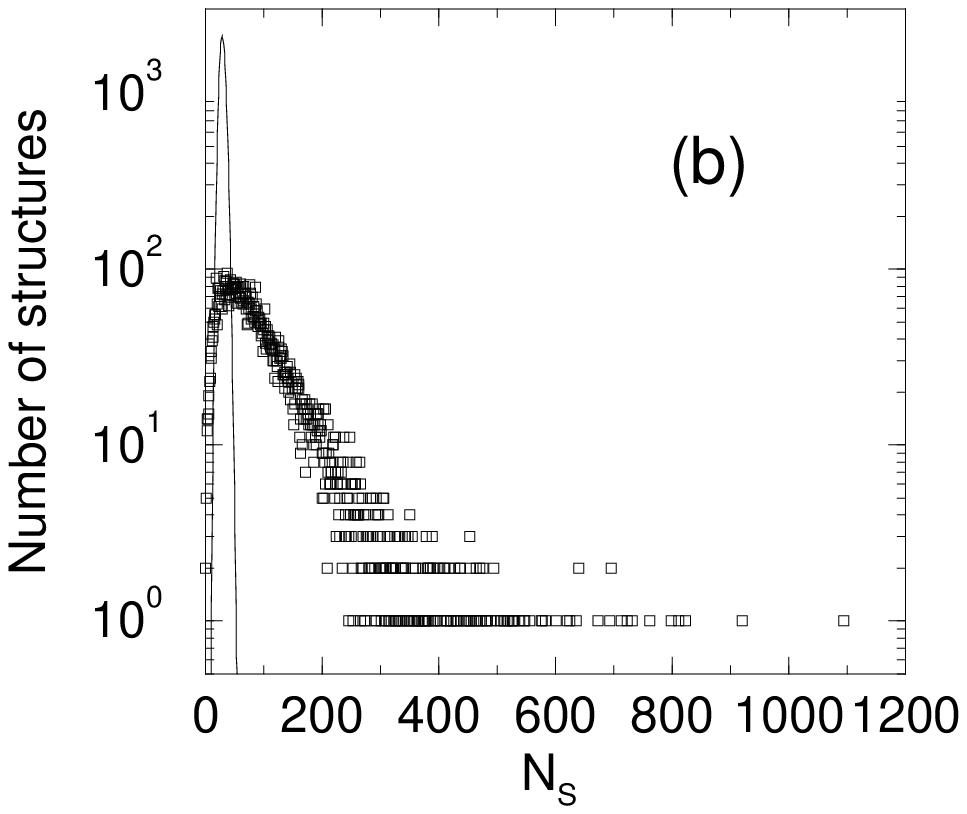,width=3in}}
\vspace{2cm}
\caption{
(a) A $6\times6$ compact structure and its corresponding string.
In the ``purely hydrophobic'' (PH) model, only two types of sites are 
considered, surface sites and core sites. The core is shown enclosed 
by a dotted line. Each structure is represented by a binary string $s_i$
$(i=1,\ldots,36)$ of 0s and 1s representing surface and core sites, 
respectively.  (b) Histogram of designability $N_S$ for the $6\times6$ 
PH model, obtained using 19,492,200 randomly chosen sequences.
A Poisson distribution with the same mean is shown for comparison.}
\label{fig1}
\end{figure}

\newpage

%Fig. 2 - (from Review Fig. 13) 
\begin{figure}
\centerline{\psfig{file=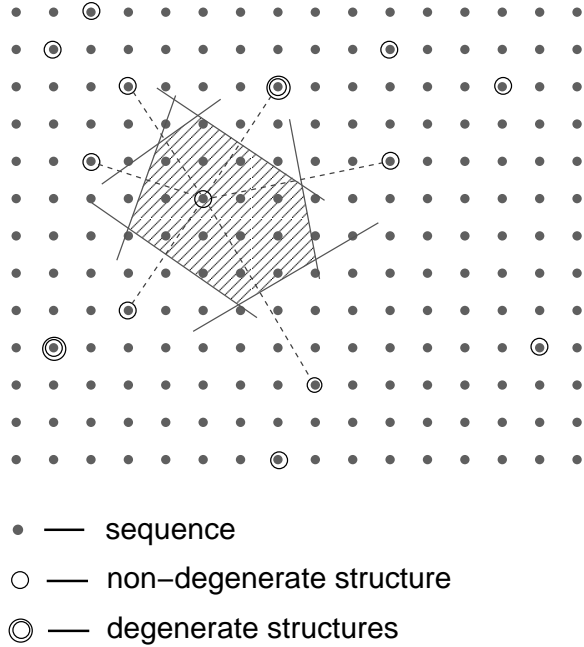,width=3in}}
\vspace{3cm}
\caption{
Schematic representation of sequences and structures in the
purely hydrophobic (PH) model. Dots represent sequences, {\it i.e.}
all binary strings. Dots with circles represent binary
strings associated with compact structures. Multiple circles indicate
degenerate strings, {\it i.e.} strings associated with more than
one compact structure. In the PH model, the energy of a sequence
folded into a particular structure is the Hamming distance between
their binary strings. Hence the number of sequences which fold
uniquely to a particular structure--the designability of the
structure--is the set of vertices lying closer to that structure
than to any other, as indicated for one particular structure by
the shaded region.}                                 
\label{fig2}
\end{figure}

\newpage

%Fig. 3 (from Review Fig. 14)
\begin{figure}
\centerline{\psfig{file=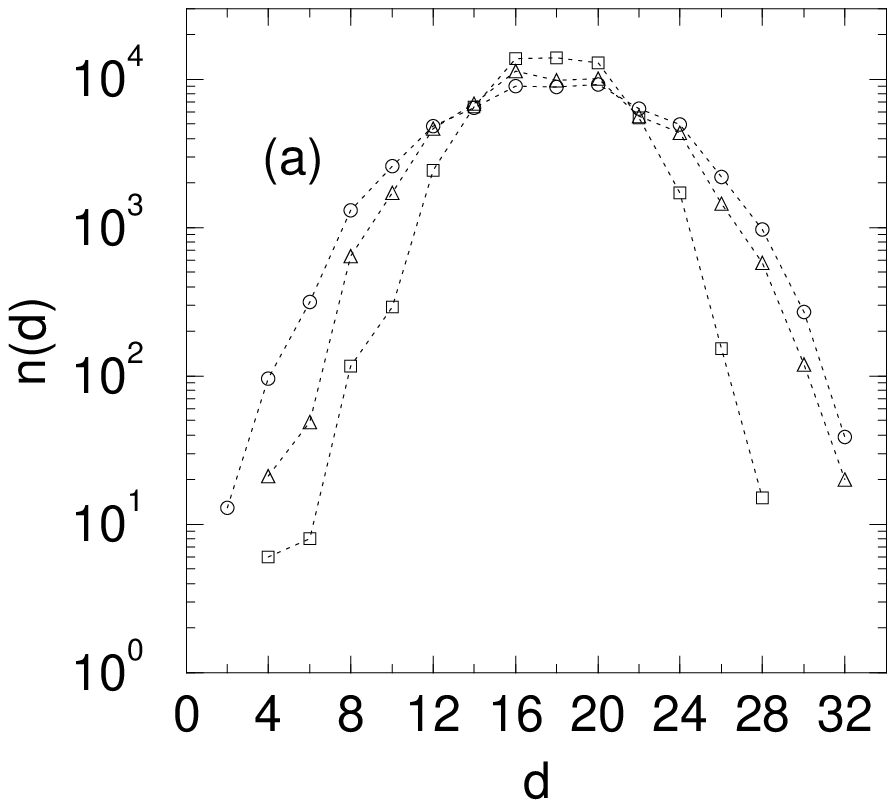,width=3in}}
\vspace{1cm}
\centerline{\psfig{file=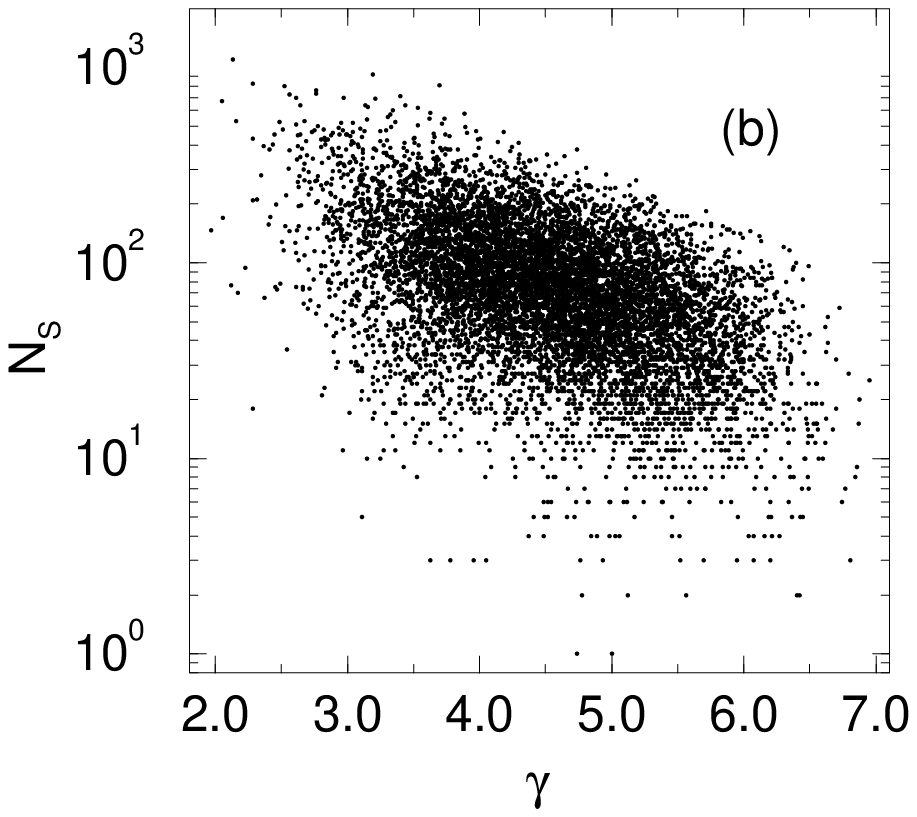,width=3in}}
\vspace{3cm}
\caption{
(a) Number of neighboring structures $n(d)$ versus distance $d$
to neighbors for three representative $6\times6$ structures, 
with low (circles), intermediate (triangles), and high (squares)
designability. The distance between structures is defined
as the Hamming distance between their binary strings.
(b) Designability versus $\gamma$, the variance of $n(d)$, 
for all $6\times6$ structures.}
\label{fig3}
\end{figure}

\newpage

% Fig. 4 (from Review 4(b)) 
\begin{figure}
\centerline{\psfig{file=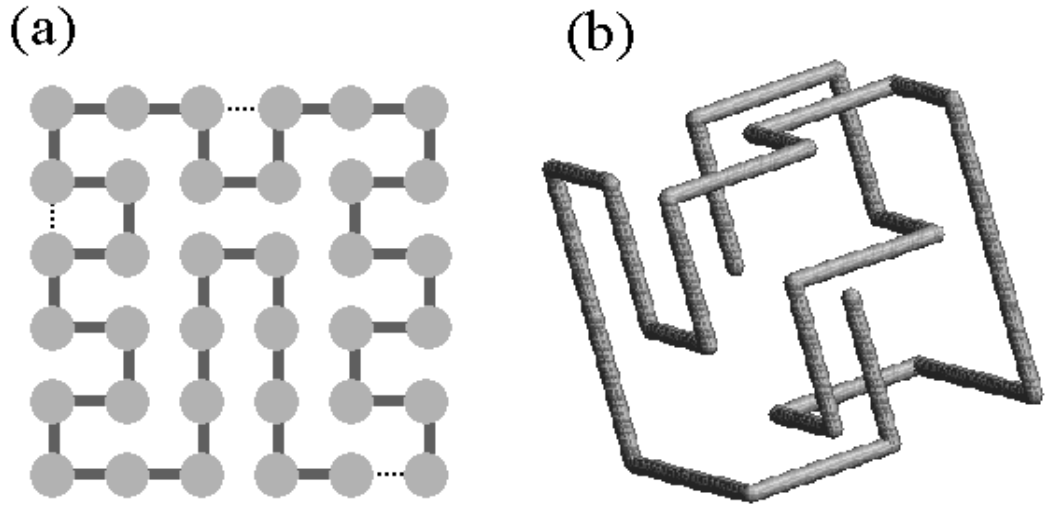,width=5.5in}}
\vspace{3cm}
\caption{
(a) Most designable $6\times6$ structure using 20
amino-acid types. Only noncovalent nearest-neighbor interactions
contribute to the energy, as indicated by dashed lines for a few pairs.
Interaction energies are
taken from the Miyazawa-Jernigan (MJ) matrix. (b) Most designable
$3\times3\times3$ structure using the same MJ-matrix energies.
}
\label{fig4}
\end{figure}                                                                 

\newpage

% Fig. 5 (from Review Fig. 9)
\begin{figure}
\centerline{\psfig{file=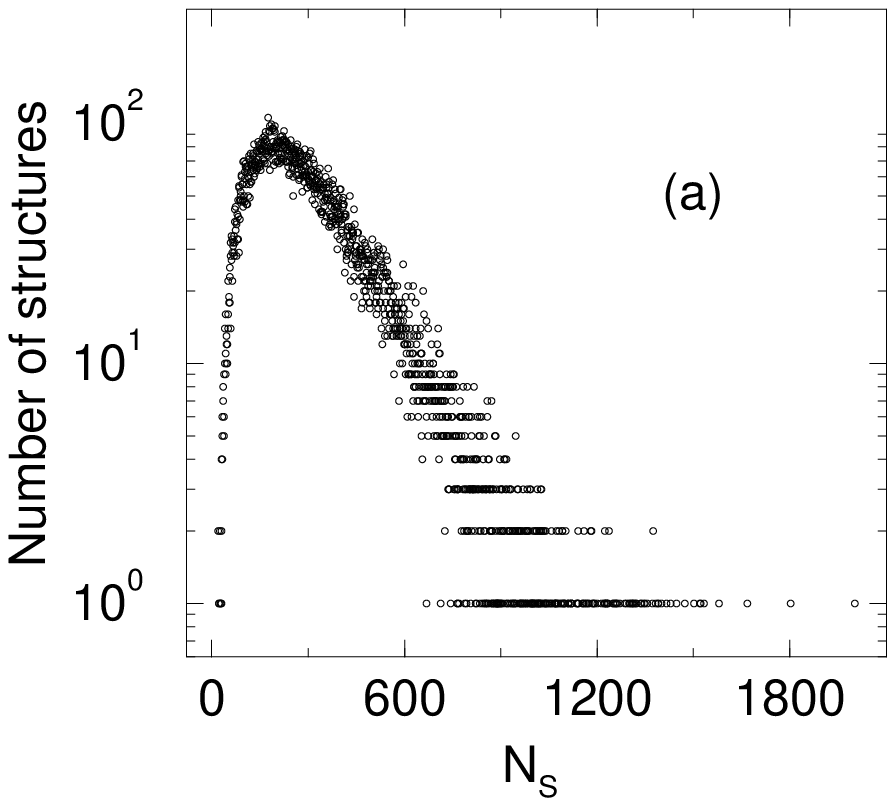,width=3in}}
\vspace{1cm}
\centerline{\psfig{file=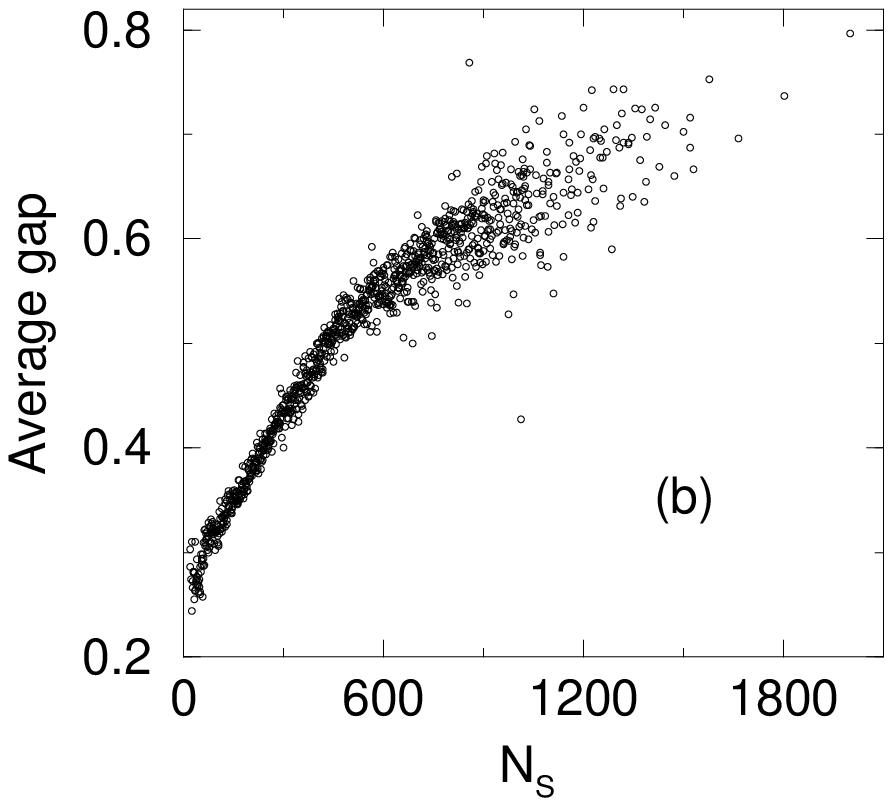,width=3in}}
\vspace{2cm}
\caption{
(a) Histogram of designability $N_S$ for the $6\times6$ MJ-matrix
model. (b) Average gap versus designability for the $6\times6$
MJ-matrix model. Data obtained using 9,095,000 randomly chosen
sequences.}
\label{fig5}
\end{figure}

\newpage

% Fig. 6 (from 20-letter paper 7(c) and 8(a)) 
\begin{figure}
\centerline{\psfig{file=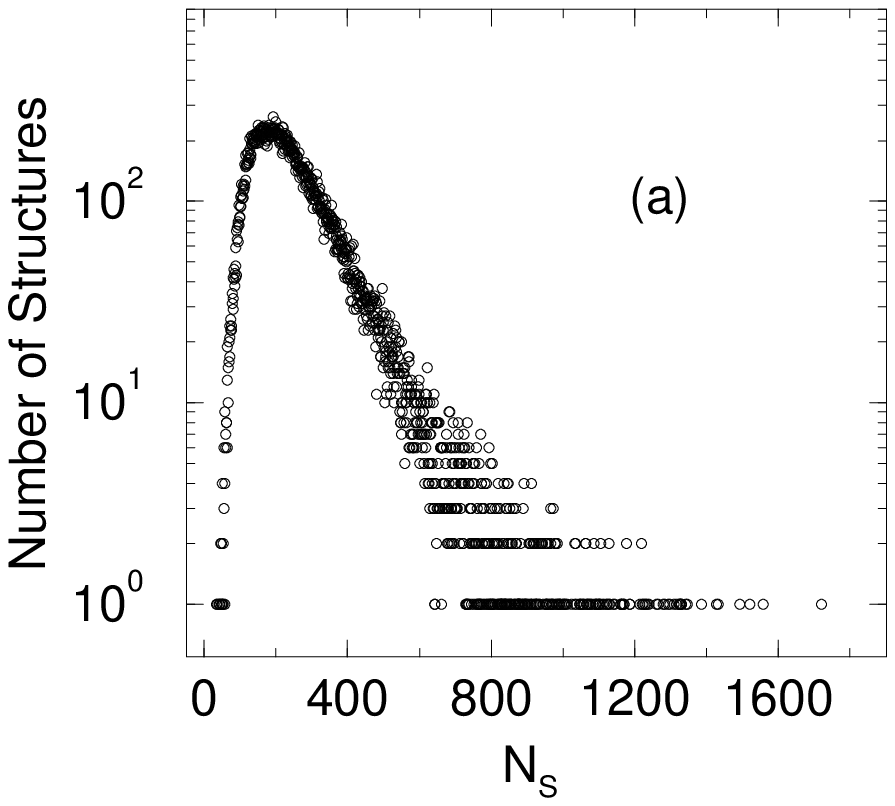,width=3in}}
\vspace{1cm}
\centerline{\psfig{file=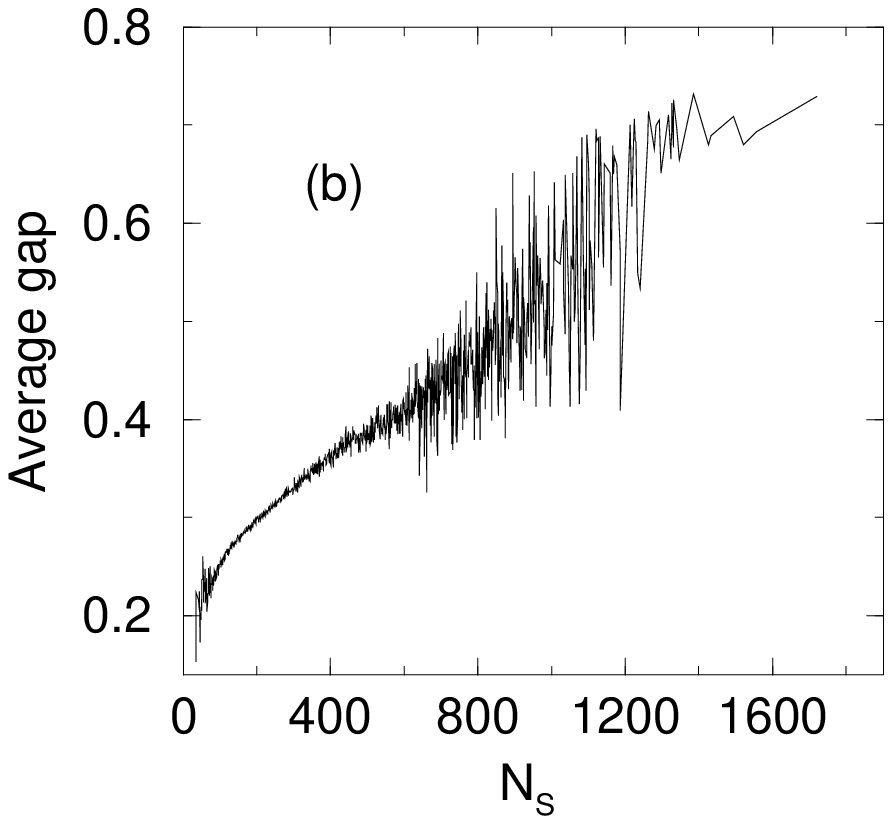,width=3in}}
\vspace{2cm}
\caption{ 
(a) Histogram of designability $N_S$ for the $3\times3\times3$ MJ-matrix
model. (b) Average gap versus designability for the 
$3\times3\times3$ MJ-matrix model. Data obtained 
using 13,550,000 randomly chosen sequences.
}
\label{fig6}
\end{figure}

\newpage

% Fig. 7  (figure from your talk)
\begin{figure}
\centerline{\psfig{file=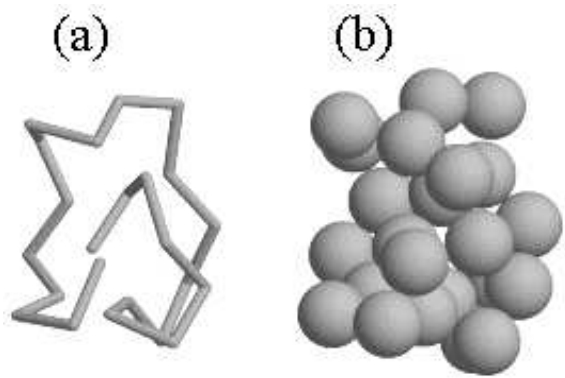,width=3in}}
\vspace{3cm}
\caption{
(a) Example of a compact, self-avoiding 23-mer backbone generated using three 
dihedral-angle pairs. (b) Backbone with generic sidechain 
spheres centered on C$_\alpha$ positions. 
}
\label{fig7}
\end{figure}

\newpage

% Fig. 8 (from Jonathan paper)
\begin{figure}
\centerline{\psfig{file=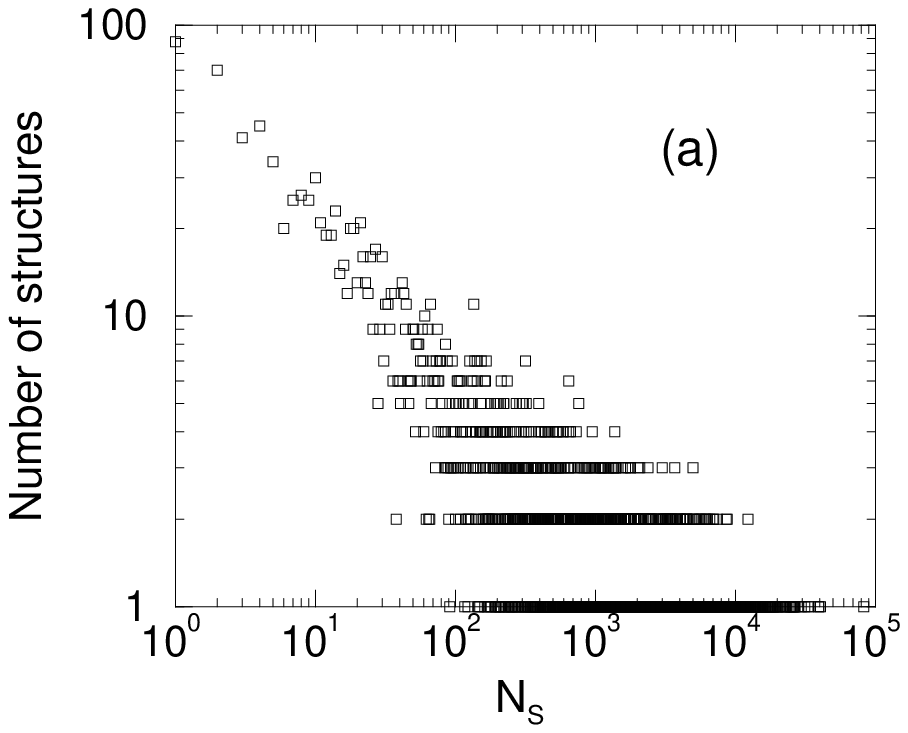,width=3in}}
\vspace{1cm}
\centerline{\psfig{file=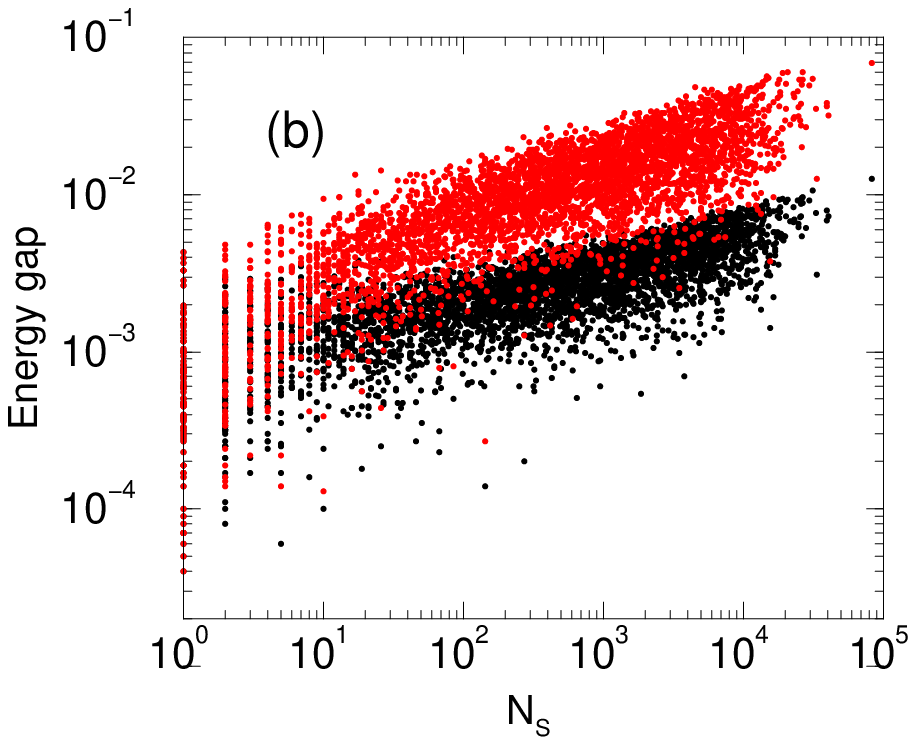,width=3in}}
\vspace{2cm}
\caption{ 
(a) Histogram of designability for 23-mer off-lattice structures
of the type shown in Fig.~\protect\ref{fig7}.
(b) Average energy gap (black dots) and largest energy gap
(red dots) versus designability. Data generated by enumeration
of all binary sequences.
}
\label{fig8}
\end{figure}

\newpage

% Fig. 9 (from Eldon packing paper Fig. 4)
\begin{figure}
\centerline{\psfig{file=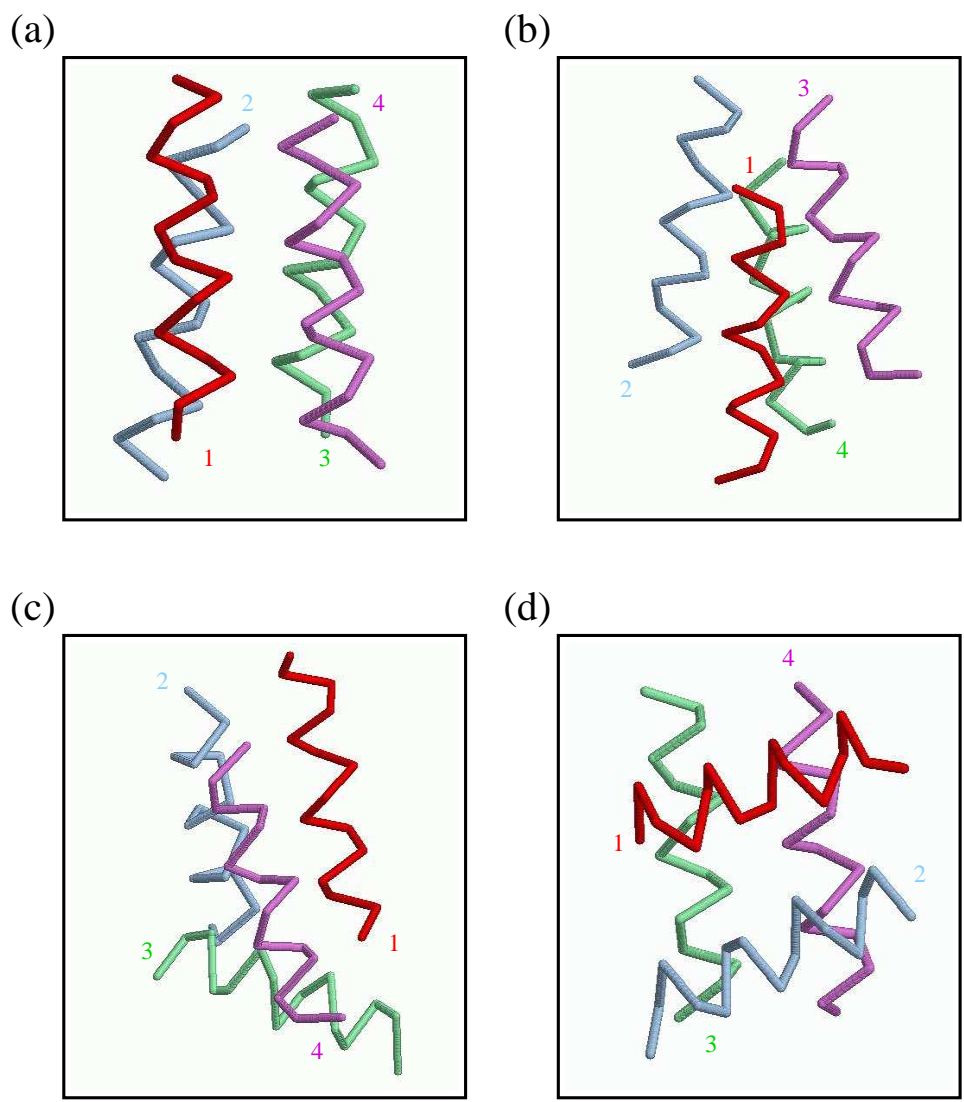,width=5.5in}}
\vspace{3cm}
\caption{ 
Four most designable four-helix bundles generated
by packing tethered 15-residue $\alpha$-helices. The helices are 
numbered at their N-terminals.}
\label{fig9}
\end{figure}

\newpage

%Fig. 10 (from Eldon packing paper Fig. 3) 
\begin{figure}
\centerline{\psfig{file=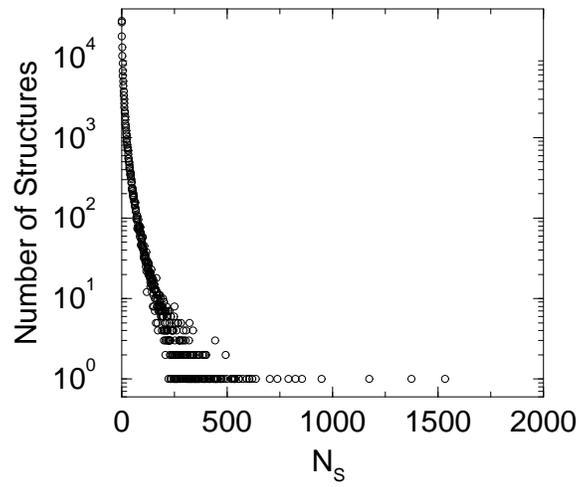,width=3in}}
\vspace{3cm}
\caption{ 
Histogram of designability $N_S$ for four-helix bundles. 
Data obtained using 2,000,000 randomly chosen binary
sequences.}
\label{fig10}                                                                 
\end{figure}                                                                 


\begin{thebibliography}{9}
\bibitem{Hamming}
The Hamming distance between two binary strings, $s_i$ and $t_i$, where
$i=1,2,\cdots,N$, is defined as $\sum_{i=1}^N |s_i-t_i|$.
\end{thebibliography}
\end{document}